\documentclass[aps,superscriptaddress,twocolumn,showkeys,showpacs]{revtex4-2}
\input{luatex-pdf}

\usepackage[T1]{fontenc} 
\usepackage{mathpazo} 
\usepackage{amsfonts}
\usepackage{braket}
\usepackage{graphicx}
\graphicspath{ {./} }
\usepackage{mathptmx}      
\usepackage{bm}
\usepackage{enumerate}
\usepackage{subfigure}
\usepackage{color}
\usepackage{array}
\usepackage{tabularx}
\usepackage{lipsum}
\usepackage{tikz}	
\usepackage{geometry}
\usepackage{qcircuit}
\usepackage{floatrow}
\usepackage{array}
\usepackage{tikzit}
\usepackage{pgfplots}
\usepackage{qcircuit}
\usepackage[toc,page]{appendix}
\usepackage{notes2bib}
\usepgfplotslibrary{groupplots,dateplot}
\usetikzlibrary{patterns,shapes.arrows}
\pgfplotsset{compat=1.13}

\tikzstyle{new style 0}=[fill=red, draw=black, shape=circle]
\tikzstyle{color}=[fill=none, draw={rgb,255: red,49; green,49; blue,49}, shape=circle]
\tikzstyle{rectangle text}=[fill=none, draw=black, shape=rectangle]
\tikzstyle{new style 1}=[fill=none, draw=black, shape=circle]

\tikzstyle{light}=[-, fill=none, draw={rgb,255: red,189; green,189; blue,189}]
\tikzstyle{arrow}=[draw=black, ->]
\tikzstyle{doublearrow}=[<->]

\usepackage{tabularx}

\usepackage{amsmath,amssymb}

\definecolor{DarkGreen}{RGB}{0,100,0}
\begin{document}
\title{Improved variational quantum eigensolver via quasi-dynamical evolution}

\author{Manpreet Singh Jattana}
\affiliation{Institute for Advanced Simulation, J{\"u}lich Supercomputing Centre, Forschungszentrum J{\"u}lich, D-52425 J{\"u}lich, Germany}
\affiliation{RWTH Aachen University, D-52062 Aachen, Germany}

\author{Fengping Jin}
\affiliation{Institute for Advanced Simulation, J{\"u}lich Supercomputing Centre, Forschungszentrum J{\"u}lich, D-52425 J{\"u}lich, Germany}

\author{Hans De Raedt}
\affiliation{Institute for Advanced Simulation, J{\"u}lich Supercomputing Centre, Forschungszentrum J{\"u}lich, D-52425 J{\"u}lich, Germany}
\affiliation{Zernike Institute for Advanced Materials,\\
University of Groningen,
NL-9747 AG Groningen, The Netherlands}

\author{Kristel Michielsen}
\affiliation{Institute for Advanced Simulation, J{\"u}lich Supercomputing Centre, Forschungszentrum J{\"u}lich, D-52425 J{\"u}lich, Germany}
\affiliation{RWTH Aachen University, D-52062 Aachen, Germany}
\affiliation{Correspondence email: k.michielsen@fz-juelich.de}
\date{\today}

\begin{abstract}
The variational quantum eigensolver (VQE) is a hybrid quantum-classical algorithm designed for current and near-term quantum devices. Despite its initial success, there is a lack of understanding involving several of its key aspects. There are problems with VQE that forbid a favourable scaling towards quantum advantage. In order to alleviate the problems, we propose and extensively test a quantum annealing inspired heuristic that supplements VQE. The improved VQE enables an efficient initial state preparation mechanism, in a recursive manner, for a quasi-dynamical unitary evolution. We conduct an in-depth scaling analysis of finding the ground state energies with increasing lattice sizes of the Heisenberg model, employing simulations of up to $40$ qubits that manipulate the complete state vector. In addition to systematically finding the ground state energy, we observe that it avoids barren plateaus, escapes local minima, and works with low-depth circuits.  For the current devices, we further propose a benchmarking toolkit using a mean-field model and test it on IBM Q devices. Realistic gate execution times estimate a longer computational time to complete the same computation on a fully functional error-free quantum computer than on a quantum computer emulator implemented on a classical computer. However, our proposal can be expected to help accurate estimations of the ground state energies beyond $50$ qubits when the complete state vector can no longer be stored on a classical computer, thus enabling quantum advantage. 

\end{abstract}

\maketitle
\section{Introduction}\label{INT}
The advent of quantum computing has seen a growing interest in developing useful applications for current and near-future quantum devices. Currently, and in the foreseeable future, quantum devices are expected to remain error-prone due to noise and decoherence. Hybrid quantum-classical algorithms, which are to some extent resilient to errors, have been proposed to integrate quantum and classical resources \cite{Moll2018, Peruzzo2014,Colless2017,Sharma2020, Wang2019, Rattew2019}. The variational quantum eigensolver (VQE) \cite{Peruzzo2014, McClean2016} and the quantum approximate optimization algorithm \cite{Farhi2014, Zhou2020, dw} are among the proposed candidates to address chemical and combinatorial optimization problems, respectively. Prototype problems have been experimentally demonstrated \cite{Kandala2017, Kokail2019, Qiang2018,Pagano2019, OMalley2016}. Simulations have also analysed such variational methods \cite{Fedorov2021, Reiner2019, Seki2020, Bharti2020}. 

Despite the progress, all demonstrated applications fall within the small-scale proof-of-concept domain. While the VQE has a simple description, there is a lack of thorough understanding when applied beyond a small number of qubits. Recent works have suggested that such explorative endeavours, which are expected to require large numbers of parameters, will encounter barren plateaus \cite{McClean2018} in the energy landscapes, diminishing all hope for quantum advantage. Furthermore, hitherto new problems in the largely unexplored large-scale simulations may be waiting for us. Thus, it is of immediate importance to investigate VQE beyond the small scale to establish a potential quantum advantage.

In this work, with a focus on the Heisenberg model, we contribute in three major aspects. First, we propose and test a general `evolution' heuristic that can, by construction, systematically lower the ground state energy. We benchmark the performance of the heuristic on one-, two-, and three-dimensional lattices, as well as several randomly generated Hamiltonians. Second, we perform large-scale simulations of VQE using an ideal quantum computer emulator and analyze the performance trend as a function of an increasing number of qubits. Last, we study the experimental realizations of our methods. We ask whether finite samples are sufficient to accurately approximate the ground state energy. We propose and test a benchmarking toolkit suitable for current and future devices, using a physically relevant problem.

The paper is structured as follows. In Sec.~\ref{secvqe}, we review the working of VQE and discuss the current problems that variational simulations face.  In Sec.~\ref{secse}, the theory of a `state evolution' heuristic is introduced which builds upon VQE.  In Sec.~\ref{secpse}, the heuristic is tested for the Heisenberg model and randomly generated Hamiltonians. In Sec.~\ref{seclsa}, we move on to large-scale applications of VQE and systematically study the performance for increasing lattice sizes. In Sec.~\ref{secla}, we discuss the relevant aspects for experimental realization, propose and test the benchmarking toolkit, and discuss if quantum advantage is feasible through VQE for the Heisenberg model.

\section{Variational quantum eigensolver\label{secvqe}}

\begin{figure}
\includegraphics{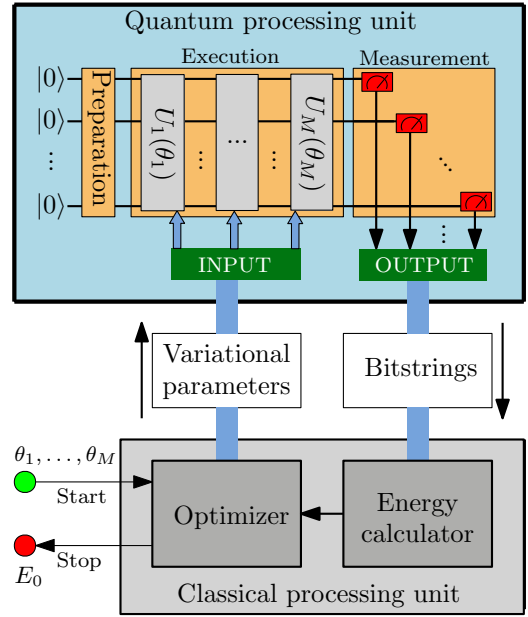}
\caption{(Colour online) Schematic of the VQE consisting of quantum and classical processing units. The circuit prepares a problem-specific initial state from $\ket{0}^{\otimes n}$ and successively applies parametrized evolutions $U_1(\theta_1), \ldots, U_M(\theta_M)$. The bitstrings from the measurement of the final state are fed to the classical unit, tasked with calculating the energy and suggesting parameters that will iteratively minimize it.  \label{figvqe}}
\end{figure}
\begin{figure}
\includegraphics[scale = 0.8]{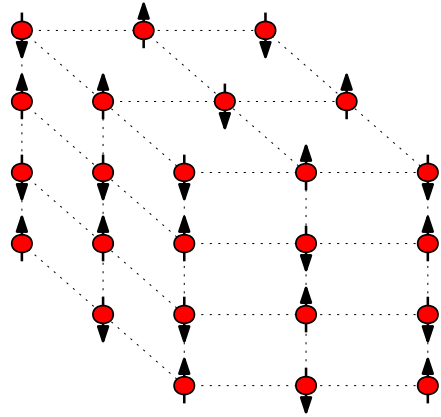}
\caption{ (Colour online) Illustrative example of a $3\times 3 \times 4$ bipartite spin lattice, where each spin is mapped to a qubit and VQE tries to find the ground state energy of the system.  \label{fig3d}}
\end{figure}

Figure~\ref{figvqe} visualizes the working of the VQE. As a hybrid method, it requires a quantum and a classical processing unit. The calculation is started by giving some initial parameters as input to the classical unit. The quantum unit prepares a problem-specific initial state and takes as input the variational parameters suitably placed in a quantum circuit that needs to be executed on the device. After the execution, a measurement gives the output as a sequence of bits whereby each bit corresponds to the measurement outcome of each qubit. The bitstring is transferred to the classical unit, which is tasked to accumulate the bitstrings and calculate the energy once a certain number of bitstrings is reached. The calculated energy is fed to a classical optimization algorithm, optimizer in short, which computes the next set of parameters with the aim that successive iterations minimize (or maximize) the energy. When some internal criteria of the optimizer are met, it stops the procedure. The last (or lowest) calculated energy is then the ground state energy.

Although the variational quantum eigensolver has been originally proposed in the context of quantum chemistry, the idea is general and readily usable for physics (and other) problems. We focus on its application to the quantum spin model given by
\begin{equation}
H = \sum_{\langle i,j\rangle}^N {J}_{ij} \bm\sigma_i\cdot \bm\sigma_j , \label{eq4:ham1}
\end{equation}
where $\langle i,j \rangle$ are pairs on a hypercube lattice of $N$ sites, $\bm\sigma \in \{\sigma^x,\sigma^y,\sigma^z\}$ are the Pauli matrices, and $J>0$. One benefit of studying spin models, in contrast to fermionic ones, is that such models find direct mappings to qubits and do not require transformations \cite{Jordan1928, Bravyi2002, Setia2018} that may yield prohibitively long circuits.

We analyze the model for simple one-, two-, and three-dimensional lattices visualized in Fig.~\ref{fig3d}. A horizontal or vertical layer illustrates a two-dimensional lattice and the edges of such a layer, a one-dimensional ring. Regardless of the dimension of the model, each spin interacts only with the nearest neighbour(s), shown through dotted lines. The alternating configuration of spins shown in Fig.~\ref{fig3d} is called the N\'{e}el state. For an even number of spins, the ground state is found in the zero-magnetization sector. The mapping of the spins to qubits is straightforward; all spins pointing up are mapped to the initial state $\ket{0}$ and those anti-parallel to $\ket{1}$. For the example shown in Fig.~\ref{fig3d}, the preparation step of the quantum processing unit in Fig.~\ref{figvqe} leaves the state $\ket{0}$ unchanged for spins pointing in one direction and flips it to $\ket{1}$ for those in the other direction.

\subsection{The variational principle}
Let $E$ be the energy of a system described by a Hamiltonian $H$, and $E_0$ its ground state energy. Then according to the variational principle, $E$ given by a parameterized wavefunction $\psi(\theta)$ is a strict upper bound to the ground state energy $E_0$,  
\begin{equation}
	E = \braket{\psi(\theta )|H|\psi(\theta ) } \geq E_0.\label{eq4:1}
\end{equation}
VQE relies on Eq.~(\ref{eq4:1}) to find the ground state energy. 

\subsection{Essential modules}

Several factors play a role in variational methods. We briefly discuss the essential modules. 

\textit{Optimizer}: Tasked with lowering the energy at successive iterations and stopping the calculation, an optimizer plays a central role. There are two broad categories of optimizers: gradient-based and gradient-free.  Depending on whether a device or an (ideal) emulator is used, each has its own merits and drawbacks. This work focuses on gradient-based optimizers since they converge faster for noise-free energy evaluations \cite{PhysRevResearch.2.043158}. We use the SLSQP and BFGS algorithms \cite{scipy, kraft, nw, opti}. For use on a quantum device, efficient (stochastic) methods have been developed \cite{Arrasmith2020, Sweke2020, OBrien2019, Mitarai2019, Stokes2020, Bergholm2018}.

\textit{Ansatz}: The choice of an ansatz is crucial to algorithms. As the current noisy intermediate-scale quantum era \cite{Preskill2018} devices can handle only low depth circuits, we focus on these. In this paper, we examine a few different ansatzes that produce low-depth circuits and require a small number of parameters. 

\textit{Initial state}: The optimization process can be significantly accelerated if an efficient initial state is known for the problem. An example is the Hartree-Fock state in quantum chemistry \cite{AIQuantum2020, Romero2019}. For our problem, it is the N\'{e}el state \cite{refId0,Kittel2004}. 

\textit{Initial parameters}: In order to take advantage of a specific initial state, the parameters in an ansatz also need to be appropriately chosen. Clever choices of initial parameters help avoid barren plateaus \cite{Rattew2019, 08572, 13751, 123123asd}. In contrast, random initialization consists of randomly choosing each parameter's value in the interval $[0,2\pi)$.

\subsection{Current problems and progress}

\begin{figure}
\includegraphics[scale = 1.0]{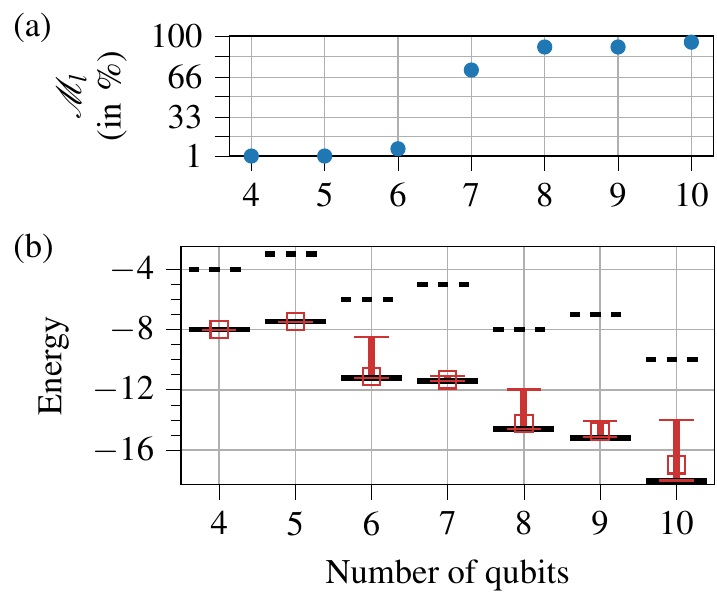}
\caption{(Colour online) (a) Percentage of unique values of energies for different problem sizes. (b) Mean distribution of the energies (red squares) for different lattice sizes, error bars are max and min values, for 100 runs. Solid (dashed) black lines are exact energies for the ground (N\'{e}el) state. \label{fig1unq}}
\end{figure}

Recent works relevant for the spin model given by Eq.~(\ref{eq4:ham1}) focus towards implementation on current devices \cite{Kandala2017,Ostaszewski2021} as well as outlining efficient schemes for similar problems \cite{Liu2019, Fujii2020}. They consider a few qubits and one-dimension only. The one-dimensional isotropic case of the model is analytically solvable using the Bethe ansatz \cite{Bethe1931}. Attempts have been made to implement VQE for this model on a quantum computer \cite{Nepomechie2020, VanDyke2021}. 

Even for small problems requiring a few qubits, the global minimum of the multi-dimensional rugged energy landscapes can be surrounded by multiple local minima. A simple numerical demonstration to show this is carried out as follows. We implement the "simple" case of one-dimensional isotropic anti-ferromagnetic rings with different numbers of spins. The BFGS optimizer is restarted $100$ times, and the parameters are assigned randomly from the interval $[0,2\pi)$. It suffices to count unique values of energy for the purpose of demonstration. We accommodate the fact that the minima may be a valley (in multi-dimensions) by rounding from floating-point precision to $10^{-3}$ for the counting. Figure~\ref{fig1unq}(a) shows the number of local minima, $\mathcal{M}_l$, per 100 trials. For larger ($N\geq 7$) problem sizes, the landscapes have multiple local minima. Figure~\ref{fig1unq}(b) shows that the average energy for 100 runs is very close to the ground state in all cases, demonstrating that when the local minima exist, they surround the global minimum. Therefore, it is vital that variational methods are equipped with a procedure to escape from local minima. Recent works in this direction are limited and resource expensive \cite{Wierichs2020}.

Despite the fact that the system memory, which doubles per qubit, for simulating up to about $30$ qubits is available on modern personal computers, medium scale simulations of the model given in Eq.~(\ref{eq4:ham1}) beyond $15$ qubits are rare in the literature. One reason for this is that the memory requirement is not the only barrier to such simulations. Large circuit depths and thousands of iterations required to reach reasonable accuracy demand considerable resources beyond the small scale. For example, Ref.~\cite{Lyu2020} reported that the simulations for the one-dimensional spin model for $20$ qubits required a week even with $192$ CPU cores. We accommodate the resource demand for our simulations on a supercomputer \cite{JUWELS} using a massively parallel emulator \cite{DeRaedt2019, DeRaedt2007}. The same emulator performed quantum verification in the supremacy experiment \cite{Arute2019}.

The phenomenon of barren plateaus \cite{McClean2018,Zhao2021,Campos2021,Patti2020,Holmes2021} is seen as a significant hurdle for increasing the problem sizes. Therefore, it is necessary to keep the number of parameters as low as possible without sacrificing accuracy, not only to avoid barren plateaus but also to keep the total computational time to a minimum. Without actually performing large-scale simulations, it remains untested whether such problems can be effectively countered. Methods to tackle all the above-mentioned obstacles are essential for leveraging the power of future quantum devices using VQE. We address some of these problems in the present work.

\section{State evolution using VQE}
\label{secse}
In this section, we propose a method that builds on VQE and systematically tries to find the ground state of a problem Hamiltonian.  We also introduce an ansatz that we will use for our simulations. 

\subsection{Evolution of the optimized state}
During the variational optimization, the wavefunction at each iteration as a function of $M$ parameters is 
\begin{equation}
\ket{\psi} = U(\bm \theta)  \ket{\Psi_0}= U_M(\theta_M)\ldots U_1(\theta_1) \ket{\Psi_0}. \label{eq1a}
\end{equation}
Let  $\mathbb{U}(\bm \Theta)$ represent the unitary operators corresponding to the optimized numeric values of the parameters. In order to avoid mixing of the optimized and unoptimized parameters, we denote the optimized parameters with $\Theta$. In our notation, the VQE performs the task $U(\bm \theta) \to \mathbb{U}(\bm \Theta)$. The state, after the optimizer signals convergence, is given by
\begin{equation}
\ket{\Psi_1} = \mathbb{U}(\bm \Theta) \ket{\Psi_0}. \label{eq1p}
\end{equation}
We propose that the final state from Eq.~(\ref{eq1p}) serves as the initial state, $\ket{\Psi_1} \to \ket{\Psi_0}$, for another variational optimization. We substitute Eq.~(\ref{eq1p}) in Eq.~(\ref{eq1a}), such that
\begin{equation}
\ket{\psi} = U(\bm \theta) \mathbb{U}(\bm \Theta)   \ket{\Psi_0}=U(\bm \theta) \ket{\Psi_1} \label{1qa}.
\end{equation}
Let the (final) wavefunction after $p^\text{{th}}$ successive repetition of the above substitution and optimization be given by
\begin{equation}
\ket{\Psi_p} =  \mathbb{U}_p(\bm \Theta_p) \ldots \mathbb{U}_1(\bm \Theta_1) \ket{\Psi_0}. \label{eqse}
\end{equation}
We call the above procedure culminating in Eq.~(\ref{eqse}) as quasi-dynamical state evolution using the VQE for each cycle, or `evolution' in short. To make use of $\ket{\Psi}$ from cycle $p$ at cycle $p+1$, the parameters in $U_{p+1}(\bm \theta)$ need to be appropriately chosen, otherwise the progress is lost. This means $U_{p+1}(\bm \theta)$ should initially be an identity circuit \cite{Grant2019}. Therefore, a suitable choice for our ansatz is $\bm \theta = [ 0,\dots,0 ]$, where by avoiding an initialization at random in the energy landscape we avoid possible barren plateaus \cite{McClean2018}.

\subsection{Choice of  $U(\bm \theta)$}
In proposing Eq.~(\ref{eqse}), it was assumed that the $U(\bm \theta)$ at each cycle is the product of the same unitary operators. This assumption can be relaxed, such that for each evolution cycle, $U(\bm \theta)$ contains either a different combination or a different set of unitary operators. In the former case, the number of parameters $M$ for all evolution cycles will remain the same. In the latter case, there are two possibilities: (a) $\geq M$ operators or (b) $\leq M$ operators. For near-term devices, option (b) is of considerable interest due to limited resources. For option (b), the maximum number of `actively' utilized parameters during each evolution cycle is $M$. Keeping the number of parameters in a circuit low has at least two benefits for our VQE simulations. First, a low number of parameters helps the classical gradient-based optimizer to converge quicker since fewer gradient computations are required per iteration. Second, the construction of the ansatz is such that the circuit depth is directly proportional to the number of parameters.  Thus, a lower number of parameters implies lower-depth circuits, and therefore usage of less computational resources. 

When building a different  $U(\bm \theta)$ at each cycle, it is currently unclear how to choose the best one for a given Hamiltonian. Adaptive methods \cite{Grimsley2019,0054822,2204.07179,PRXQuantum.2.020310,12231232} to build each operator one by one solve this problem but are computationally expensive. Due to the absence of a cost-effective way to ascertain a new $U(\bm \theta)$ for each cycle, we use the same one for each cycle. As our numerical investigations show in the following sections, this simple approach has significant benefits. However, even without the numerical evidence, Eq.~(\ref{1qa}) tells us that the number of active parameters during VQE can be carefully kept below a threshold and for each new cycle of evolution different unitary operators can be made accessible. This allows for expanding the parameter space at a minimal expense of a polynomial increase in the circuit depths. 

In Ref.~\cite{ Skolik2021}, a similar approach with the difference that parameters added in each cycle can be optimized together with the previous parameters was suggested, and demonstrated for an $8$ qubits circuit only. It has been shown to be useful for combinatorial optimization problems \cite{Liu2021}. Another similar technique involves changing the Hamiltonian at each iteration \cite{doi:10.1021/acs.jctc.9b01084}.

\subsection{Connection to quantum annealing}

We conjecture that for suitable choices of $\mathbb{U}(\bm \Theta)$ at each cycle, the evolution method facilitates finding the ground state $\ket{\Psi_g}$ of a Hamiltonian $H$, such that
\begin{equation}
\ket{\Psi_p} \to \ket{\Psi_g} \quad \text{ as } \quad p \to \infty. \label{eqconj}
\end{equation}
If finding the ground state energy $E_0$ of $H$ is the primary interest, then using Eq.~(\ref{eqconj}) and the variational principle, 
\begin{equation}
E_p \to E_0 \quad \text{ as } \quad p \to \infty. \label{eqen}
\end{equation}
Thus, repeated applications of evolution cycles will systematically lower the energy, under the condition that suitable $\mathbb{U}(\bm \Theta)$ are chosen for each cycle. The proposal remains a conjecture because it is unclear what the suitable $\mathbb{U}(\bm \Theta)$ are for each cycle and given problem Hamiltonian.

The adiabatic theorem \cite{Born1928} guarantees that a system, (say) initially prepared in the ground state, will remain in its instantaneous ground state if the change in the system Hamiltonian is slow enough and if there is a sufficiently large gap between the ground and excited states. Since a change in the system Hamiltonian is equivalent to a change in the unitary operators as given by Eq.~(\ref{eqse}), the theorem guarantees that Eq.~(\ref{eqconj}) holds. By application of the variational principle on Eq.~(\ref{eqconj}), it is further guaranteed that Eq.~(\ref{eqen}) holds. Since an optimizer in the VQE algorithm is designed only to accept those parameters that ultimately lower the energy, the construction of the evolution heuristic guarantees that the energy is either lowered or stays constant. For $p\to \infty$, one can imagine choosing the parameters $\bm \Theta$ such that the state evolution corresponds to an adiabatic evolution. A similar line of thought has helped to develop the quantum approximate optimization algorithm \cite{Farhi2014} and rapidly quenched quantum annealing \cite{Callison2021}.

From a numerical simulation perspective, the small or large but ultimately finite value of $p$ depends on the computational resources required for each cycle. Our numerical evidence shows that indeed by increasing $p$, there is a systematic decrease in energy. We discuss these results in the following sections. Our numerical evidence suggests that the evolution method is a useful heuristic.

\subsection{Our ansatz}
The choice of an ansatz is crucial not only from a computational perspective but also for the evolution heuristic. If an ansatz can accurately represent the ground state energy of a given problem, optimization of the parameters may still pose problems, and the evolution heuristic can help overcome them. On the other hand, if an ansatz has a poor overlap with the ground state, the heuristic can still help achieve quasi-optimal results. The ansatz we consider approximates the ground state energy reasonably well, but not exactly, for the larger problems of interest \cite{jattana2022assessment}. The ansatz is given by 
\begin{equation}
	U(\bm \theta) = \prod_{\substack{l = N-1\\ k=N}}^{\substack{l = 1\\ k = l+1}} U_{lk}(\theta_{lk})  \prod_{\substack{l = N-1\\ k=N}}^{\substack{l = 1\\ k = l+1}} U_{kl}(\theta_{kl}), \label{eq4:xy1}
\end{equation}
where 
\begin{equation}
	U_{kl}(\theta_{kl}) =
	\begin{cases}
		e^{-i\theta_{kl} \sigma_k^y\sigma_l^x } & \text{if $k=N$ or $l=N$,}\\
		e^{-i\theta_{kl} \sigma_k^y\sigma_l^x\sigma_N^z } & \text{otherwise.}\label{eqxy1}
	\end{cases}   
\end{equation}
The number of unitary operators in this ansatz is given by $N(N-1)$. We note that, as pointed out previously in Refs. \cite{Grimsley2020, Tranter2019, Izmaylov2020}, the ordering of the factors in Eq.~(\ref{eq4:xy1}) is important. Any other ordering can produce results that may be different from one another. We name the specific combination of the factors in Eq.~(\ref{eq4:xy1}) as the XY-ansatz. The operators defined in Eq.~(\ref{eqxy1}) can be easily implemented in terms of single- and two-qubit gates on a quantum circuit. The details about the implementation are given in Appendix~\ref{app5}. 

The choice of the XY-ansatz is motivated by three empirical considerations. First, the ansatz produces low-depth circuits that can be implemented on near-term quantum devices.  Second, the parametrized gate is always placed on only one qubit. Due to manufacturing imperfections, not all qubits in a device perform equally well, and the XY-ansatz allows the best performing qubit to implement the parametrized gates. For current devices having higher two-qubit gate error rates, the best performing qubit may be the best connected qubit. Third, the ansatz has an \textit{all-to-one} connectivity between the qubits, which is easier to implement in devices, in contrast to an all-to-all connectivity \cite{Skolik2021}.

\section{Performance of the state evolution method}
\label{secpse}
We compare the performance of our methods to the strategy of random initialization of the parameters. The random initializations (RI) strategy involves assigning random values to the parameters in the range $[0,2\pi)$. We present results for the Heisenberg spin model and several randomly generated problem Hamiltonians.

\subsection{Heisenberg model}

\begin{figure}
\includegraphics[scale = 1.0]{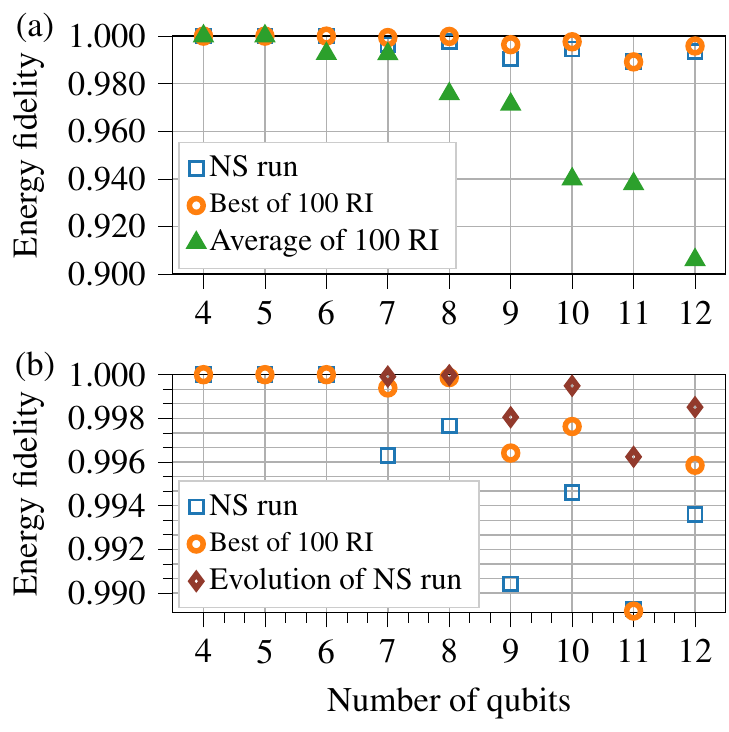}
\caption{(Colour online) (a) Comparison between the N\'{e}el state (NS) and random initializations (RI) approaches for different numbers of qubits for the anti-ferromagnetic rings. The energy fidelity is defined as $E^f/E_0$. (b) Same as (a) except that the data for the average of RI runs has been replaced by the evolution data and that the scale is different. \label{fig4a}}
\end{figure}

We test our heuristic on one-dimensional anti-ferromagnetic rings of length $4\leq N \leq 12$. Three types of strategies are used, namely, RI; initialization from the N\'{e}el state (NS) where the parameters are initialized as zeros; and the evolution heuristic. We calculate the energy fidelity given by the ratio of the final energy obtained using VQE and the ground state energy. As shown in Fig.~\ref{fig4a}(a), for $N\leq6$, NS performs just as well as the best out of $100$ RI optimization runs -- both can find the ground state energy. The average performance of RI, however, begins to deteriorate for $N\geq 6$. Keeping the RI runs to $100$, we observe a significant drop in the average performance as $N$ increases. The drop in the performance can be understood from the perspective that as the number of parameters is increased, a larger number of restarts would be required to keep up the performance. Meanwhile,  if the best RI run is better than NS, it is only slightly so. On the other hand, the initial energy for NS is lower than RI, requiring fewer energy evaluations until optimization convergence (data not shown). Thus, NS achieves similar accuracy by reducing the huge computational cost from $100$ RI runs to just one run.

We perform our evolution heuristic on the final state at the end of the optimization of NS. In this way, we leverage the $\ket{\psi_1}$ obtained after the optimizer gets stuck in a local minimum for each $N\geq 7$. We know that these are local minima because RI runs find a lower energy. For each evolution cycle, we use the same $U(\bm \theta)$, and the stopping criterion is when the change in energy is less than $10^{-4}$. As initial parameters we use zeros for each new cycle, the reason for which is discussed in Appendix~\ref{appyy}. The results are shown in Fig.~\ref{fig4a}(b). The odd-spins lattices have relatively lower energy fidelities due to the degeneracy of the ground state. Our heuristic further increases the energy fidelities in all cases, beyond the best of RI runs and significantly better than the average of RI runs. The evolution heuristic successfully overcomes several local minima for all $N$. One way to interpret how it overcomes is as follows. Once an optimizer gets stuck, a new cycle is started which can be seen as another VQE process but with a new initial state. Such a configuration change creates a restructuring of the energy landscape where a larger or different part of the Hilbert space becomes accessible. The energy obtained from the initial parameters loaded in the new cycle is no longer located at a local minimum as in the previous cycle. Using the evolution heuristic, we get the triple benefit of having a smaller parameter space, low-depth circuit, as well as a higher energy fidelity.

\subsection{Random Hamiltonians}
\begin{figure}
\includegraphics[scale = 1.0]{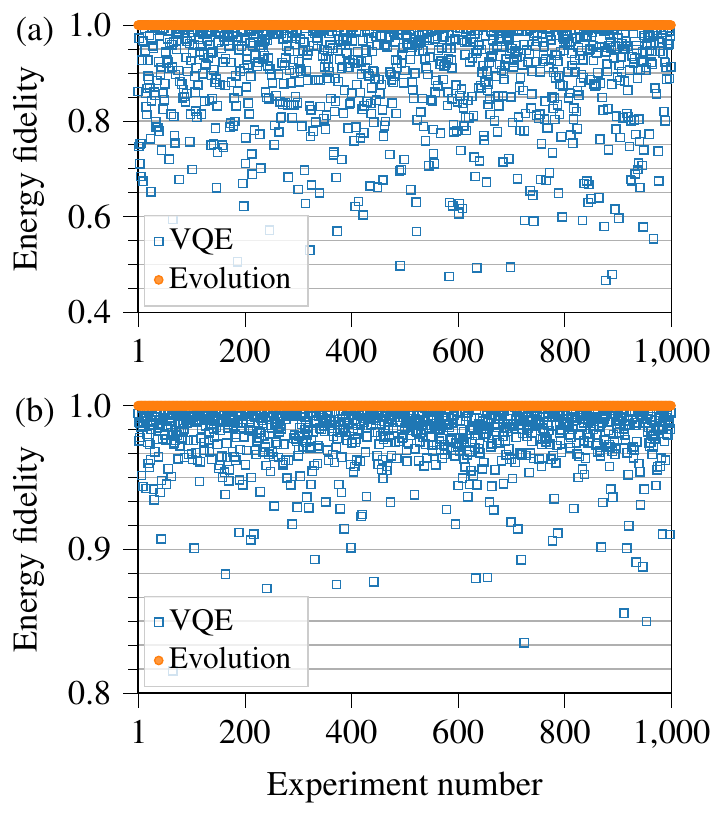}
\caption{(Colour online) (a) Ratio of the ground state energies as obtained by the evolution heuristic for a thousand randomly generated problem $N=6$ instances. Blue squares are the minimum energies found using VQE, and orange circles the improved energies using evolution. The energy fidelity is defined as $E^{(1)}/E^{(2)}$. (b) Same as (a) except for $N=8$.\label{figh}}
\end{figure}

We study the performance of the evolution heuristic by studying many randomly generated Hamiltonians. To generate the problems,  we define the Hamiltonian
\begin{equation}
H = \sum_{\langle i,j\rangle}^N \Big( {J}_{ij}^{xx} \sigma_i^x \cdot \sigma_j^x+{J}_{ij}^{yy} \sigma_i^y \cdot \sigma_j^y+{J}_{ij}^{zz} \sigma_i^z \cdot \sigma_j^z \Big),\label{eq4:hammf}
\end{equation}
where the sum $\langle i,j\rangle$ sums over all pairs of $N$ lattice sites. We randomly select one-third of all the terms in $H$. To the selected terms, we assign random values to all the coefficients ${J}_{ij}^{ \alpha\alpha } \in (0, 10)$ for $\alpha\in   \{x,y,z\}$. While Eq.~(\ref{eq4:hammf}) is analytically solvable for ${J}_{ij}^{xx} = {J}_{ij}^{yy} = {J}_{ij}^{zz} = 1$, the energy for a finite subset with random coefficients can only be computed numerically. 

We generate two sets of random Hamiltonians, specifically for $N=6$ and $N=8$. In the case of $N=6$, we use only half the number of operators in the XY-ansatz, which in itself produces a different ansatz. Since the XY-ansatz can be split into a product of two groups of unitary operators $U_2(\bm \theta_2) U_1(\bm \theta_1) \ket{\Psi_0}$, we use only $U_1(\bm \theta_1)$ as the ansatz. We prepare the system in the N\'{e}el state. We initialize the parameters randomly for each experiment and perform the evolution on the state obtained. Each evolution cycle uses the same $U(\bm \theta)$. In Fig.~\ref{figh}(a), we show the results for the $N=6$ case. For each of the $1000$ experiments, we performed $10$ calculations per experiment such that after the first VQE calculation there are nine evolution cycles. We  plot the ratio of the final energy $E^{(1)}$ obtained using VQE (blue squares) and the final energy $E^{(2)}$ calculated (orange dots) after the evolution cycles performed on the final state obtained from VQE. In the case of $N=8$, we use the XY-ansatz, and all other settings are the same as in the $N=6$ case.  The results are shown in Fig.~\ref{figh}(b). From the results, we observe that the heuristic also works for $N=8$. We observe that there is a significant improvement in the energy by using the evolution heuristic.

The fact that  $E^{(1)}/E^{(2)}< 1$ in a large sample of random experiments for both test cases shows that the heuristic is useful in further increasing the energy fidelity without requiring a larger parameter space. While the evolution heuristic does not guarantee to find the global optimum, i.e. a fidelity of $1.0$ (data not shown), the results from this section show that it can produce quasi-optimal energy fidelities that would otherwise require a large number of randomly initialized optimization runs. Furthermore, the simple approach of using the same $U(\bm \theta)$ for each evolution cycle appears useful for a relatively large set of problems.

\section{Large-scale applications}
\label{seclsa}
We present results for large-scale applications of the evolution heuristic to find the ground state of the isotropic anti-ferromagnetic Heisenberg Hamiltonians on one-, two-, and three-dimensional lattices. In all cases, we use the XY-ansatz. We use the N\'{e}el state as the initial state and initialize the parameters to zeros. For each evolution cycle we use the same $U(\bm \theta)$.

\begin{figure}
\includegraphics[scale = 1.0]{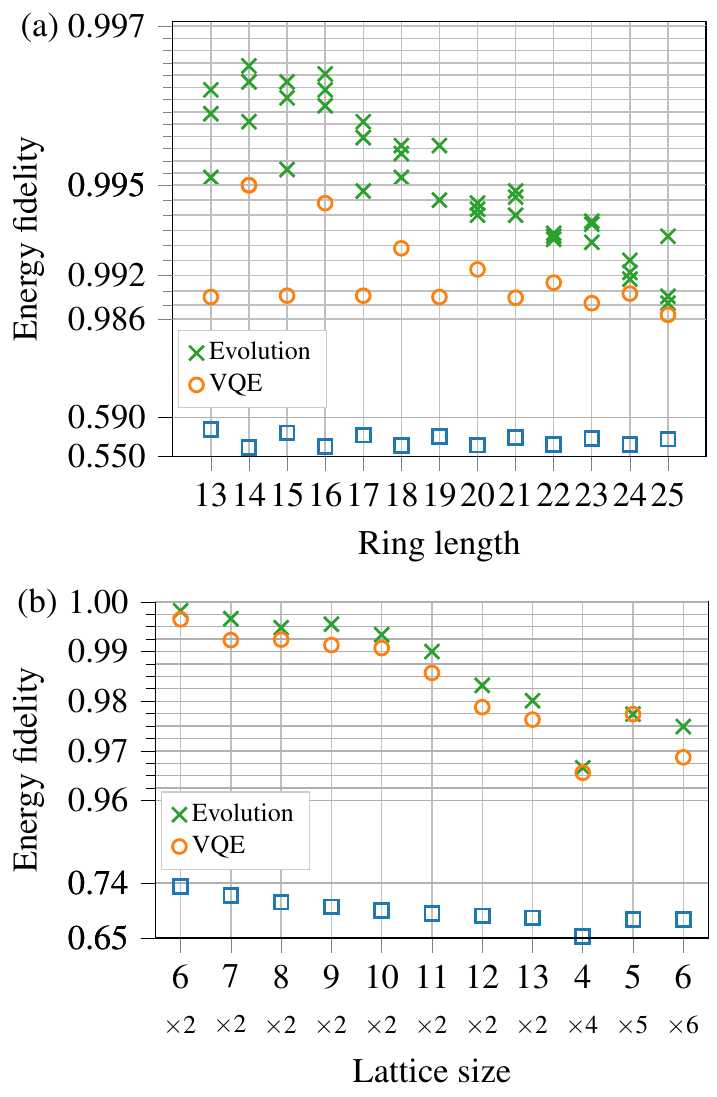}
\caption{(Colour online) (a) Evolution heuristic applied to one-dimensional anti-ferromagnetic rings. Blue squares show the energy fidelity of the N\'{e}el state, orange circles of the VQE, and green crosses of the evolution heuristic. Each cross represents the energy obtained after one cycle of evolution. (b) Same as (a) except for two-dimensional lattices of dimensions indicated on the x-axis. \label{fig5}}
\end{figure}

\subsection{One- and two-dimensional lattices}

In Fig.~\ref{fig5}(a), we show the evolution heuristic applied to one-dimensional rings of length $N>12$. For the purpose of demonstration, we perform only a few evolution cycles per lattice, given by the green crosses. Each cycle is observed to increase the energy fidelity indicated by the crosses moving higher up. Similarly, Fig.~\ref{fig5}(b) shows the results for ladder and square lattices of dimensions up to $13\times 2$ and $6\times 6$, respectively. Except for $6\times 6$, all two-dimensional lattices were chosen to have open boundary conditions. The energy per spin for the largest ladder and square lattices, after one evolution cycle, were $-2.216$ and $-2.647$ as compared to the ground state energy per spin $-2.261$ and $-2.715$, obtained by the Lanczos method \cite{cullum2002lanczos}. Note that $J=1$ and the Hamiltonian has dimensionless units. In all large-scale simulations, the low-depth XY-ansatz can find the ground state energy with a fidelity greater than $96\%$.  We observed that while a single evolution cycle increases the energy fidelity, for the two-dimensional lattices, the second cycle did not produce any variation in the energy. Furthermore, the convergence of the evolution heuristic to the ground state appears to slow down as the ring size increases. A possible reason for both observations is the breakdown of the assumption that the same $U(\bm \theta)$ is a good candidate for each cycle, suggesting that some other ansatz should be tested in the future.

The largest system that we simulated was a $40$-qubit isotropic ladder. Since the time required to perform one energy calculation was large, we removed some of the parameters. In order to remove the parameters, we rewrite the two product terms in Eq.~(\ref{eq4:xy1}) as $U_2(\bm \theta_2) U_1(\bm \theta_1) \ket{\Psi_0}$, and consider only $U_1$, which contains then only half of the parameters. Further, we roughly followed the rule $l+10\geq k>l$  and reduced the parameters from $1560$ originally to a more manageable $400$. After four iterations, the energy per spin was $-1.966$ (without evolution), which was significantly lower than the N\'eel state energy $-1.5$. The ground and first excited state energies are $-2.312$ and $-2.262$, respectively, found using the Lanczos algorithm. The VQE energy estimate is considerably higher than the ground state energy, which can be explained as follows. Firstly, restricted by the computational time required, the estimate uses only the standard VQE. Lastly, the number of parameters in the XY-ansatz was heavily reduced, resulting in reduced performance.

We also studied the evolution heuristic for the $40$-qubit ladder. We reduced the parameters even further with the rule $k=l+1$ (using $U_1$ only) to $40$. Let us denote the resulting product of operators as $U_{40}(\bm \theta)\ket{\Psi_0}$. We performed two evolutions after the initial standard VQE energy of $-1.867$ with the same $U_{40}(\bm \theta)$ per cycle, with a few iterations only, to obtain the energy per spin $-1.878$. The evolution heuristic can lower the energy even for a very small parameter space. Due to the construction of the heuristic, a different $U(\bm \theta)$ may prove helpful in further improving the energy estimate. This is a task for future work.
 
\subsection{Three-dimensional lattices}
By simulating the one- and two-dimensional lattices, we observe that the XY-ansatz appears to yield a reasonable approximation to the ground state. Similarly, for the three-dimensional lattices, we continue by initializing parameters as zeros and the system in the N\'{e}el state, and use the XY-ansatz. Since the maximum number of qubits that can be simulated on classical hardware is severely restricted by the memory of the supercomputer, only a limited number of three dimensional lattices can be studied. We simulate $3\times3\times b$ lattices where $b=2,3,4$ with open boundary conditions and $J=1$.

In Appendix~\ref{se1} we show that, as expected, the N\'eel state continues to be an efficient initial state also for the three-dimensional lattices. The N\'{e}el state gives a low initial energy per spin to start the optimization, e.g. $-1.83, -2.00,$ and $-2.08$ for $b=2,3,$ and $4$, respectively. We performed two evolution cycles for $b=2,3$. For $b=4$, the circuit contains $1260$ parameters and due to a time limit of 24 hours per run on the supercomputer, only a limited number of iterations are possible. Due to the amount of computational resources needed, we did not perform an evolution for $b=4$. During two evolutions, the energies continued to drop for $b=2$ and $3$. The final energies per spin were $-2.482, -2.631,$ and $-2.694$ which can be compared to the ground state energies $-2.617, -2.720,$ and $-2.797$ obtained by the Lanczos method.

\section{Considerations for experimental realizations} 
\label{secla}
There are several practical aspects for implementing the variational algorithms depending on the technological maturity of the quantum computing devices. In this section, we discuss relevant considerations for experimental implementations. Although the evolution framework is general and applicable to various problems, we restrict our focus to spin models of the type presented in the previous sections. 

\subsection{Hardware optimal benchmarking}

\begin{figure}
\includegraphics[scale = 1.0]{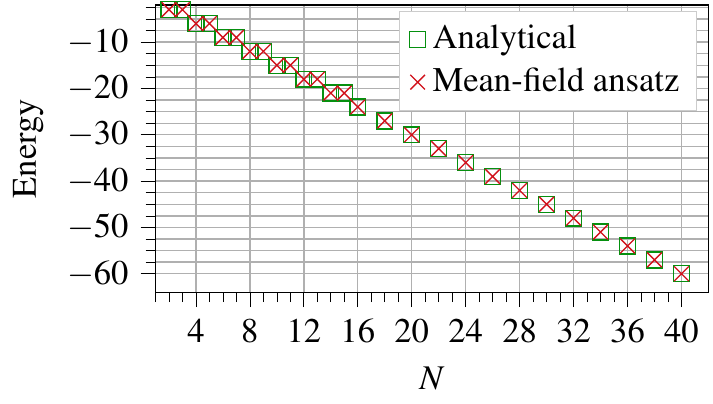}
\caption{(Colour online) Comparison of the energies obtained using the mean-field ansatz for $N$ and the exact analytical values. \label{figmf} }
\end{figure}

Benchmarking protocols are employed to ascertain the proper working of any quantum device. Most protocols involve physically irrelevant tasks, although the hope for future quantum computers is to solve physically relevant problems, e.g. in physics or chemistry.  The Heisenberg Hamiltonian could be a candidate. However, the XY-ansatz we presented requires all-to-one connectivity between the qubits. Although all-to-all connectivity is native to the trapped-ion devices \cite{Nam2020}, the number of functional qubits remains low. Alternatively, current superconducting devices, among others, which mostly offer neither all-to-all nor all-to-one connectivity but other benefits, are also dominant in the research community. Considering the limited qubit connectivities of the current devices, we propose that the mean-field Hamiltonian, given by Eq.~(\ref{eq4:hammf}) with ${J}_{ij}^{ \alpha\alpha } =1$ for all $\alpha\in   \{x,y,z\}$, is a suitable candidate for benchmarking. The ground state energy of the Hamiltonian is given by a simple and easy to verify \cite{Hams2000} expression $3(a- N)/2$ where $a = 1$ for odd $N$ and $a = 0$ for even $N$.

We propose a very low-depth `mean-field ansatz' well-suited for the mean-field Hamiltonian.  With a constant circuit depth of five, independent of the number of qubits, the mean-field ansatz accurately recovers the ground state of the model using only up to $N/2$ parameters. The details of the mean-field ansatz are given in Appendix~\ref{ap1}. In Fig.~\ref{figmf}, we show numerical evidence that the ansatz can recover the ground state energy for $N \leq 40$. Since the ground state energy for the model is the same for $N^{\text{even}}$ and $N^{\text{even}}+1$ spins, for an odd number of spins we only consider $N\leq 15$. The solution is easily verified by setting all parameters $\theta_i = \pi/2$ for $i = 1,\ldots, N/2$. Therefore, in general, the mean-field ansatz and Hamiltonians are simple benchmarks for physics problems on the current and future quantum devices. 

\onecolumngrid

\begin{figure}
\includegraphics[scale = 0.62]{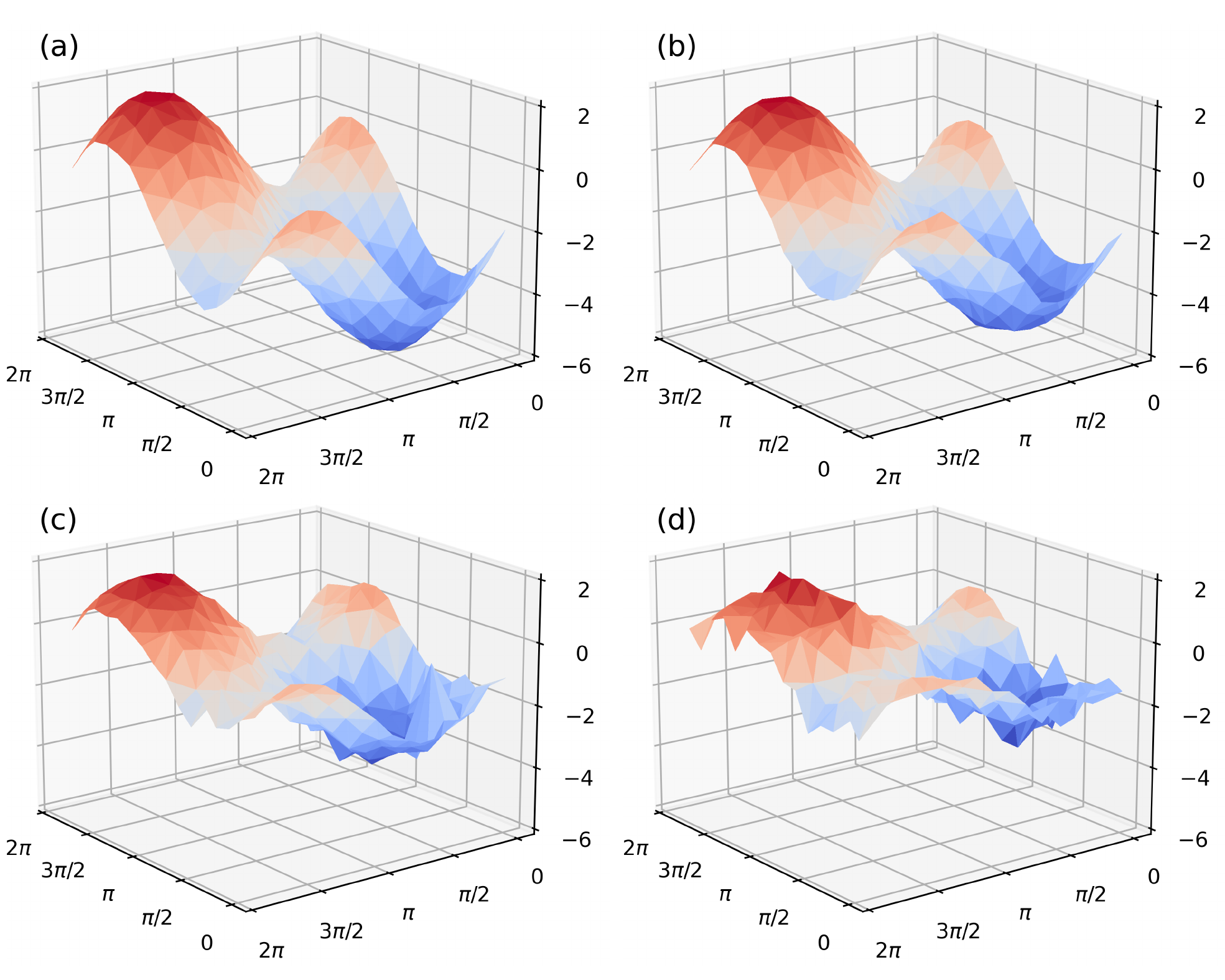}
\caption{(Colour online) Energy landscapes for the  $N=5$ mean-field Hamiltonian measured using (a) an ideal emulator, and from the IBM Q devices (b) \textit{santiago}, (c) \textit{belem} without using swap gates, and (d) \textit{belem} using swap gates. The x and y axes show the parameter values and the z-axis shows the energy scale. \label{figibm}}
\end{figure}

\twocolumngrid

Using the ansatz, we perform benchmarking of the IBM Q devices. In Fig.~\ref{figibm}(a), we show the energy landscape for the $N=5$ mean-field Hamiltonian by using a two-parameter mean-field ansatz on an ideal emulator. We create a $16 \times 16$ grid and plot the energy on the z-axis. We observe that the landscape is periodic for each parameter in the interval $\theta \in [0,2\pi]$. The global minimum is located at $\theta_1 = \theta_2 = \pi/2$. We compare the ideal landscape with those obtained from two different IBM Q devices, \textit{Santiago} \cite{santiago} and \textit{Belem} \cite{belem}, having \emph{quantum volumes} $32$ and $16$, respectively. The results are shown in Figs.~\ref{figibm}(b-d). Due to the limited connectivity between the qubits in the IBM Q devices, a careful selection of the layout is essential. We demonstrate the effects of using different layouts by choosing a layout where two adjacent qubits in our circuit are physically separated by one qubit in the middle, thus requiring swap gates. Using $2^{13}$ samples, we observe qualitative agreement to the ideal landscape for all cases. For all the tested devices, the minimum also appears to be located at $\theta_1 = \theta_2 = \pi/2$. The ground state energy, calculated by exact diagonalization, is $-6$. The minimum energies found on the devices were $-5.30$ (\textit{Santiago}), $-3.97$ (\textit{Belem} without using swap), and $-3.38$ (\textit{Belem} using swap). The quantitative difference can be attributed to the errors in the devices. We observe that the device with a larger quantum volume shows better results for our benchmark. The IBM Q Belem device performs worse when swap gates are required for the implementation, even for swapping only two qubits. Knowledge of the actual device connectivity is essential to obtain better results. The two parameters and an energy landscape having a unique global minimum reachable from all initial points makes this a simple benchmarking problem for the $N=5$ case. 

\subsection{Computational states}

\begin{figure}
\includegraphics[scale = 1.0]{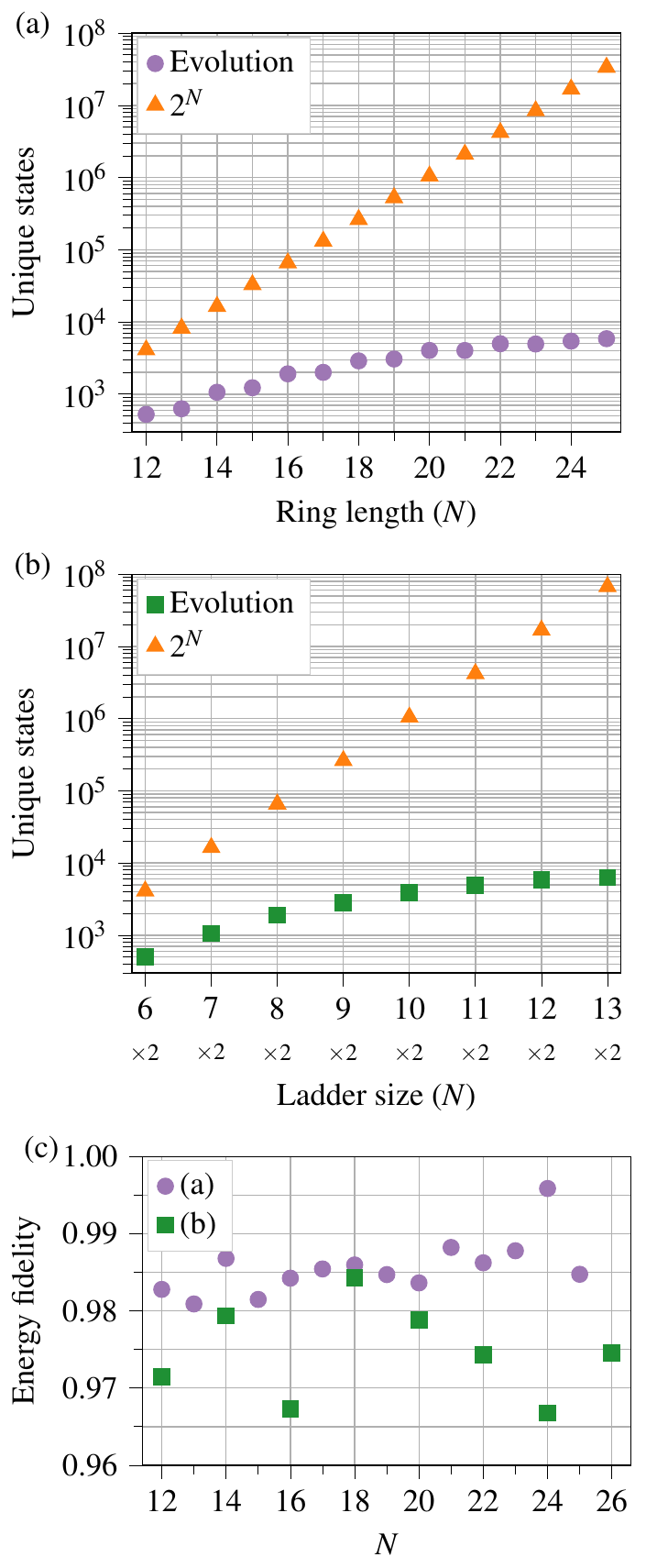}
\caption{(Colour online) (a) Comparison between the unique states generated from $\ket{\Psi}$ using $2^{13}$ samples and the size of the Hilbert space for different lengths of the one-dimensional isotropic rings. (b) Same as (a), except for ladder lattices. (c) Energy fidelity of the energies calculated from the $2^{13} $ samples.\label{fig12} }
\end{figure}

In our simulations, we performed calculations on the state vector. However, on an actual quantum device, $\ket{\Psi}$ is not directly accessible. Instead, one draws samples from the underlying distribution. The potential advantage of quantum computing rests on two closely related assumptions about the sampling. First, the number of contributing states in a certain basis (i.e. the computational basis)  from the entire Hilbert space does not increase exponentially with $N$. Second, a finite number of samples is sufficient to represent the entire distribution accurately. The hope of gaining quantum advantage is to find algorithms that fulfil the above two assumptions. This section discusses whether these two assumptions are viable for the Heisenberg model using the XY-ansatz.

From the final state  $\ket{\Psi}$ calculated on the emulator, we sample $2^{13}$ times from the underlying probability distribution.  This sample size is motivated by currently available IBM Q devices, which offer a maximum of $2^{13}$ samples per experiment. It should be noted that $2^{13}$ samples are not necessarily sufficient to accurately extract all the possible unique states for the larger lattice sizes. However, accurate energy fidelity has overriding importance. In Fig.~\ref{fig12}(a), we count the number of unique states or distinguishable bitstrings sampled from $\ket{\Psi}$ for the one-dimensional isotropic rings and compare it to the number of unique states in the Hilbert space. We perform the same task for the ladder lattices, and the results are shown in Fig.~\ref{fig12}(b). From both plots, we observe that the first assumption appears to be satisfied; that is, the number of unique states (purple dots and green squares) do not seem to grow exponentially with $N$. 

In Fig.~\ref{fig12}(c), we plot the energy fidelity obtained by calculating the energy from the $2^{13}$ samples for both rings and ladders. The fidelities can be directly compared to those obtained by calculating the energy from $\ket{\Psi}$, shown in Figs.~\ref{fig5}(a-b). We observe that the fidelities in Fig.~\ref{fig12}(c) and  Figs.~\ref{fig5}(a-b) differ by about one percent for the rings and less than three percent for the ladder lattices. The difference is further reduced by using more samples. The energy calculated from a small number of samples is reasonably close to the actual energy. Therefore, a finite number of samples is sufficient to accurately represent the distribution.

\subsection{Quantum vs classical}
\label{secqa}

Quantum advantage concerns the idea that certain tasks can be completed significantly faster on a quantum computer than on a classical computer. In this section, we compare the expected computational time required by a fully functional future quantum computer against an emulation of an ideal quantum computer using a massively parallel simulator run on a supercomputer for the task of finding the ground state of the Heisenberg model.  We question if quantum advantage is possible using VQE for the Heisenberg model. Although emulation of VQE might not be the best classical method for finding the ground state energy of the Heisenberg model on a classical computer, we restrict ourselves to the same method for the purpose of comparison.

The largest number of successive operations on a quantum computer are given by
\begin{equation}
N_o = N_g\times aT_h\times N_s, \label{eqpp}
\end{equation}
where $N_g$ is the maximum number of gates that need to be sequentially executed \cite{note1}, $T_h$ is the number of terms in the Hamiltonian to be measured, $a$ is a constant that encodes the number of groups of the Hamiltonians which can be simultaneously evaluated, and $N_s$ is the number of samples required to reach a certain statistical accuracy. The time taken to complete $N_o$ operations can be estimated by knowing the single- and two-qubit gate times. On the IBM Q superconducting type of quantum computer, single-qubit gates take about $85$ ns and two-qubit gates take about $400$ ns \cite{Takita2017}. Ion traps offer even slower two-qubit gates of the order $1.6 \mu s$ \cite{Schafer2018}. For our calculations, we take the IBM Q values.

For the $N=24$ spins ring, we count the total number of energy evaluations required up to three evolution cycles to be $84744$. At each cycle, a new layer of $U(\bm \theta)$ is added to the circuit. The largest number of single- and two-qubit sequential operations executed are approximately $6400$ and $11500$, respectively. By taking optimistic values $aT_h =3$ and $N_s=2^{13}$, the time required to calculate the energy once is about $126$ seconds. The total time required to complete the variational calculation using a quantum computer stands at $124$ days. In comparison, running the emulator on a supercomputer required $4$ days for the same computation. 

Similarly, for the $6\times 6$ and $3\times 3\times 3$ lattices, the time required to calculate the energy once is $90$ and $52$ seconds, respectively. The total number of energy evaluations were $12700$ and $68426$. The total time required to complete the variational calculation using a quantum computer stands at about $13$ and $41$ days, compared to $7$ and $4$ days on the supercomputer, respectively. Counterintuitively, the VQE runtime for the $6\times 6=36$ spins lattice is less than that of the $24$ spins ring due to the total energy evaluations and the number of evolutions performed.

The above calculations reveal that the large number of energy evaluations required to obtain the ground state energy forbids quantum computers to run VQE in a reasonable amount of time. However, as shown in Appendix~\ref{ap2}, we observed quick drops in the energy initially during the optimization which relatively slows down as it approaches a local minimum. Hence, the first few iterations provide the largest decrease in energy. For the $3\times3 \times 3$ lattice, energy per spin of $-2.605$ is found within $5\times 10^3$ energy evaluations. This will expectedly take $24$ hours on a quantum computer. 

In summary, VQE on an emulator using a classical computer performs the task faster than a hypothetical fully functional quantum computer up to the number of qubits and tasks we tested. This does not mean that emulators will always perform tasks faster than actual quantum computers. For several parameters and lattice topologies, e.g. random couplings or frustrated structures, the Heisenberg model is a hard problem. The memory requirements to store the complete state vector for a $50$ or more qubits system is beyond the current memory storage capacities of modern supercomputers. If the favourable scaling of the ansatz continues, a future quantum computer may be able to approximate the ground state energy of the Heisenberg model beyond $50$ qubits. For such large problems, the polynomial time scaling given in Eq.~\ref{eqpp} can be expected to hold. In this sense, the realization of potential quantum advantage using VQE for the Heisenberg model may be anticipated on a quantum computer.

\section{Conclusion}
We studied the performance and the experimental implementations of the variational quantum eigensolver applied to the Heisenberg model. We introduced and extensively tested a state evolution heuristic to overcome the obstacles faced by current VQE algorithms. Without increasing the active number of parameters, our tests showed that the heuristic could escape local minima. In contrast, the current standard method, which uses restarts of sets of random parameters, would quickly become intractable even for quantum computers. We observed that the evolution heuristic improves the estimates of the ground state energies of the Heisenberg model compared to the standard VQE and this across all lattice dimensions. The simulations were accelerated by initializing the quantum system in the N\'eel state, which continued to be an initial state with significantly lower energy to start the variational optimization, as compared to random initializations. Taking into account that the current and near-term devices will only have a limited computational capacity, we proposed as a benchmark a mean-field ansatz requiring a circuit depth of five independent of the number of qubits. The benchmark was able to accurately recover the ground state of the mean-field spin model using only $N/2$ parameters.  Finally, realistic gate execution time calculations of the current quantum hardware capabilities show that an emulator can perform the task of finding the ground state energy of the Heisenberg model faster than current (ideal) quantum computers. However, there is hope for quantum advantage because a computation that cannot be performed on classical hardware due to memory limitations may be performed on future quantum computers. 

Even though our heuristic works very well, innovatory improvements can benefit the evolution method's performance and cogency. One possible direction is to investigate analytical or cost-effective numerical ways to find the best $U(\bm \theta)$ for each cycle and given problem. While the variational methods are not affected by the sign problem \cite{DeRaedt1993}, it remains to be seen if the optimization algorithms will remain efficient for $N>40$. The parallelizability of VQE is another aspect that can be explored \cite{2111.05176}. The evolution heuristic can be seen as complementary to the idea of optimising all parameters actively, i.e. the entire parameter space can be made active when evolution stops making progress and vice versa. Another critical issue is the connectivity of the XY-ansatz; while the heuristic is ansatz-independent, the development of a hardware-connectivity efficient ansatz that also accurately approximates the ground state of the Heisenberg model is an open problem. The effect of quantum noise should also be studied in future works. While we have shown that the simulations find the ground state for up to the limit of what can be classically simulated using a limited number of samples, it remains to be seen if this favourable scaling continues beyond what can be classically simulated. Our results indicate that it can be expected.

\section{Acknowledgement}
The authors gratefully acknowledge the Gauss Centre for Supercomputing e.V. (www.gauss-centre.eu) for funding this project by providing computing time \cite{JUWELS}. We acknowledge use of the IBM Q for this work. The views expressed are those of the authors and do not reflect the official policy or position of IBM or the IBM Q team. M.S.J.  acknowledges support from the project OpenSuperQ (820363) of the EU Quantum Flagship.

\begin{appendices}
	
\section{The XY-ansatz}
\label{app5}

\begin{figure}
	\includegraphics[scale = 0.07]{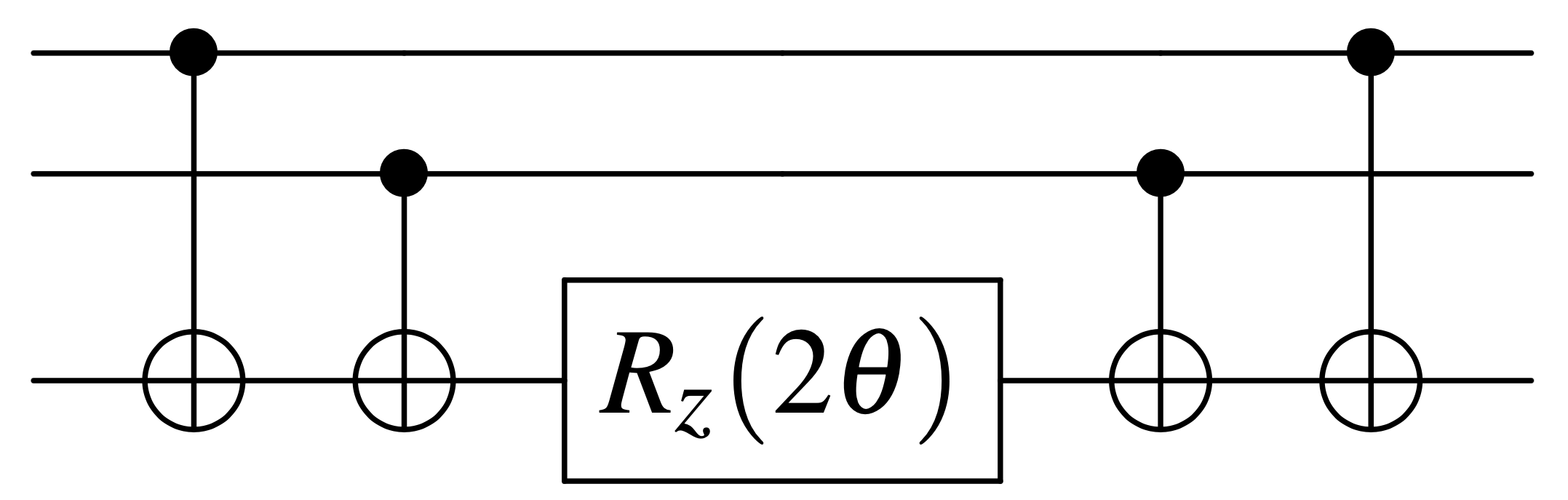}
	\caption{Circuit showing the implementation of $\text{exp}(-i \theta \sigma^Z\otimes \sigma^Z\otimes \sigma^Z)$.}\label{fig4:zzterm}
\end{figure}

We illustrate how the XY-ansatz is implemented as a quantum circuit. The  ansatz consists of products of unitary operators of the forms $\text{exp}(-i\theta A\otimes B)$ and $\text{exp}(-i\theta A\otimes B\otimes C)$, which would require factorization into products of unitary operators \cite{DeRaedt1987} and significantly increase the circuit depth. Instead, when $A=B=C = \sigma^z $, we take the implementation \cite{chaung} given by the circuit shown in Fig.~\ref{fig4:zzterm}. To implement $ A,B \in \{ \sigma^x,\sigma^y \}$ as in the case of the XY-ansatz, we appropriately change the basis as done in \cite{OMalley2016}. When the number of qubits is larger than two, after changing the basis, we always place the parametrized gate on the qubit with the largest index. After several trials we found that by using operators of type  $\text{exp}(-i\theta A\otimes B\otimes C)$ to place all the parametrized gates on one qubit, the performance was consistently better. We believe that this aspect of quantum circuit preparation needs further exploration, but this is outside the scope of the current work.

\section{Evolution parameters}
\label{appyy}
We discuss numerical examples that shed light on what parameters need to be employed when starting a new evolution cycle. We study two possibilities: all parameters initialized either randomly or as zeros.

\begin{figure}
\includegraphics[scale = 1.0]{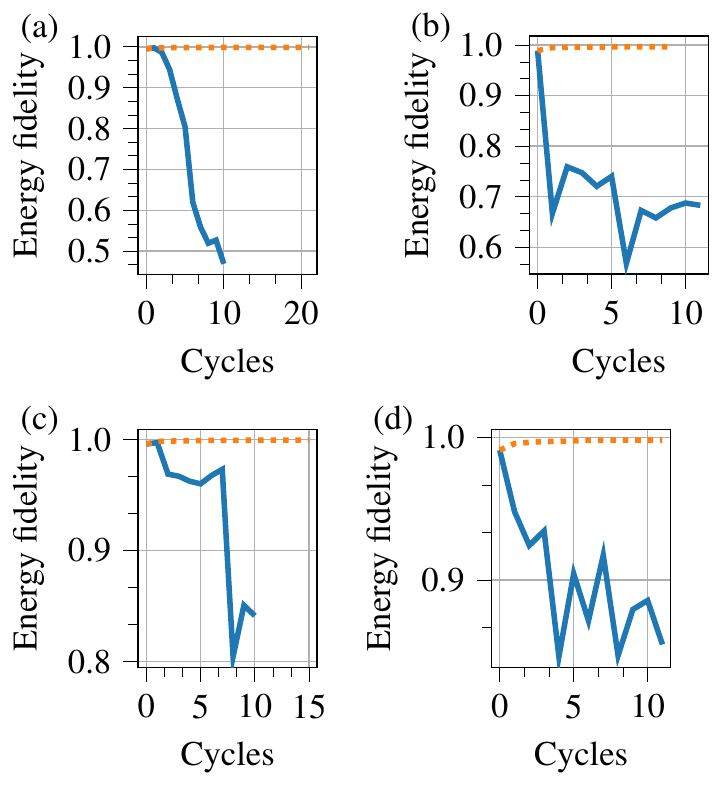}
      \caption{(Colour online) Comparison of the change in energy fidelity when using random parameters (solid blue) and zeros (dotted orange) for the initial parameters of each new evolution cycle. (a) 12 qubits; (b) 11 qubits; (c) 10 qubits; and (d) 9 qubits. \label{figsss}}
\end{figure}

As shown in Fig.~\ref{figsss}, the energy fidelity drops rapidly if random new non-zero parameters are used instead of zeros. While starting a new evolution cycle with parameters as zeros increases the energy fidelity (not visible in the scale on plot), random parameters show the opposite effect. This is understood from Eq.~(\ref{1qa}), where $\ket{\psi}$ can be used effectively in the next evolution cycle only if for the initial parameters $\bm \theta$
\begin{equation}
U(\bm \theta) \mathbb{U}(\Theta)\ket{\Psi_0} = \mathbb{U}(\Theta)\ket{\Psi_0}. \label{llk}
\end{equation}
An evolution cycle that systematically lowers the energy at each iteration can be achieved by setting $U(\bm \theta = [0, \ldots, 0])$ in Eq.~(\ref{llk}), but not by setting the parameters randomly. Additionally, setting parameters as zeros is equivalent to an identity circuit and is shown to help avoid barren plateaus \cite{Grant2019} (see Appendix~\ref{se1}). By using random parameters,
\begin{equation}
	U(\bm \theta) \mathbb{U}(\Theta) \neq \mathbb{U}(\Theta),
\end{equation}
and $\ket{\Psi}$ is not preserved for the next cycle.

\section{mean-field ansatz\label{ap1}}
\begin{figure}
\caption{Circuit implementation using the mean-field ansatz for $N=6$. }\label{figq}
\includegraphics[scale = .12]{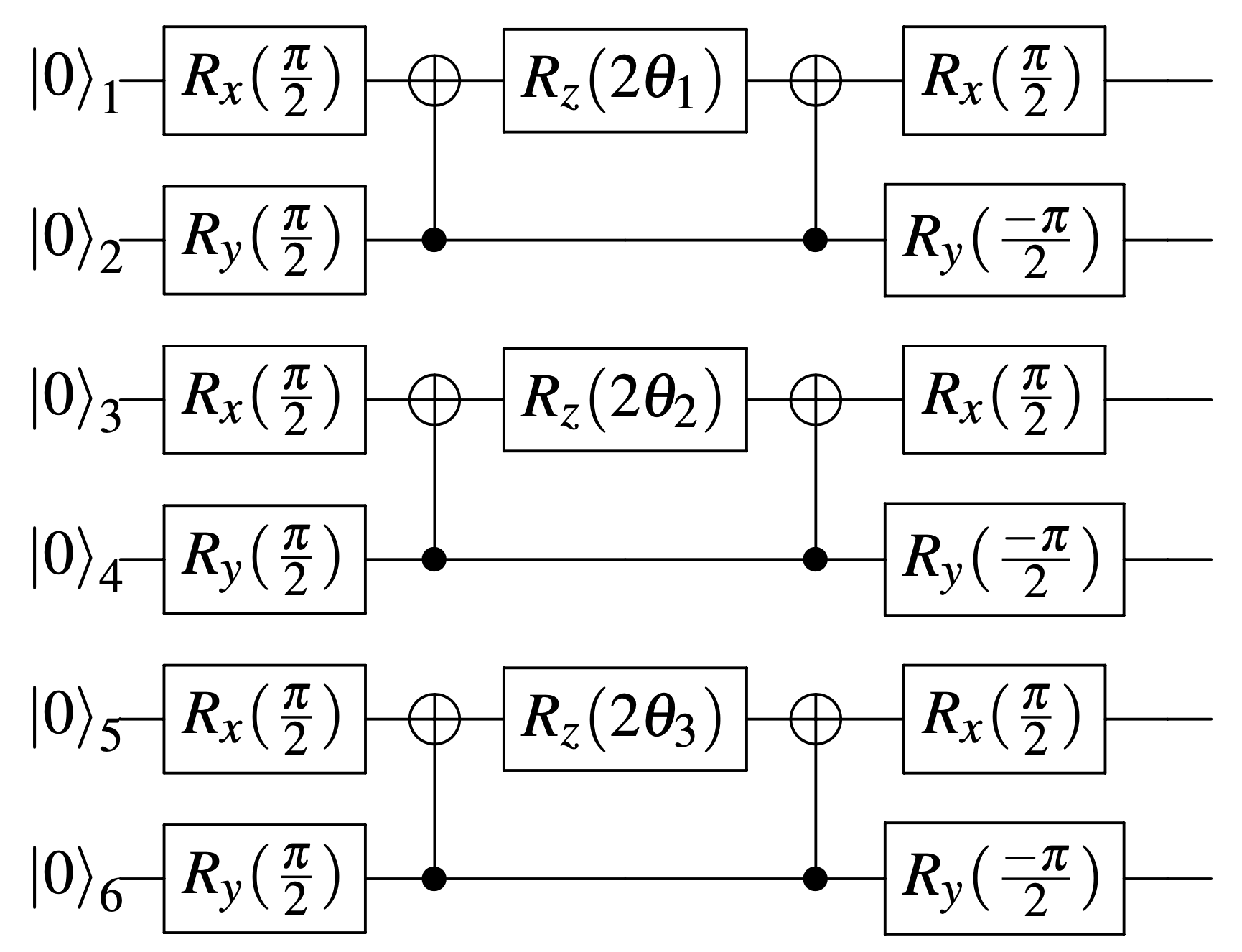}
\end{figure}

The ansatz is given by 
\begin{equation}
	U(\bm \theta) = \prod_k e^{-i\theta_k \sigma^x_k \sigma^y_{k+1}  },
\end{equation}
where $k = 1,3,5,\ldots< N$ and the initial state is 
\begin{equation}
 \ket{\Psi_0} = \ket{\ldots 0101},
\end{equation}
where the odd and even indexed (starting at $0,1,2,\ldots$) qubits are initialized in the state $\ket{0}$ and $\ket{1}$, respectively. We construct an optimized circuit for the mean-field ansatz as shown in Fig.~\ref{figq}.

The ansatz only requires a nearest neighbour connectivity for quantum devices, which is readily available for current devices. For the $N=5$ case discussed in the main text, we construct the same ansatz as one would for a four-qubit circuit and leave the fifth qubit without any gates. Due to the simplicity of the problem, such an approach appears to work. The final state appears to be a product of two-qubit states. Once the ansatz is constructed, the mean-field Hamiltonian terms can be measured by rotating to the appropriate basis. 

\section{Initial state efficiency}
\label{se1}
\begin{figure}
	\includegraphics[scale = 0.78]{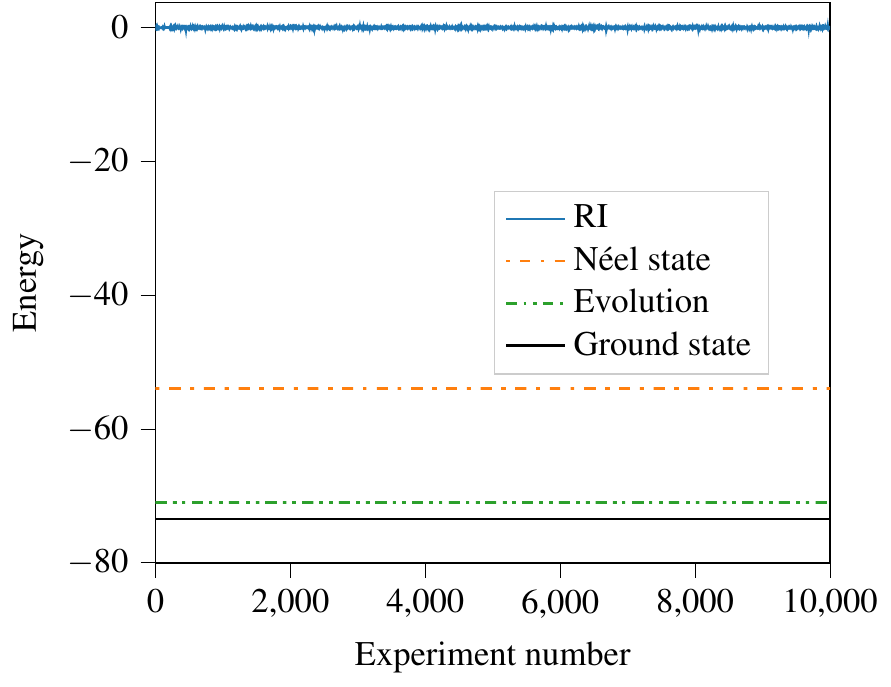}
	\caption{Comparison of the energies obtained using random initial parameters (RI) and the N\'eel state for $10^4$ random initializations. The N\'eel state was used in our evolution heuristic to obtain an approximate ground state energy. \label{figbba}}
\end{figure}

We compare the initial energy obtained using the N\'eel state and using randomly generated parameters (RI). As a specific example, we consider the $3\times 3\times 3$ lattice where the circuit contains $702$ variational parameters. We choose $10^4$ different random sets of values for the parameters and plot the energy obtained in each case. We do not perform a VQE calculation but plot the energy corresponding to the initial parameters. Figure~\ref{figbba} compares the energy obtained using RI runs and the N\'eel state. Given that $10^4$ random points in the energy landscape correspond to $E \approx 0$, the landscape is relatively flat or barren across a vast region. We expect that the gradients computed around the points where the energy was calculated would be close to zero in this case. This is numerical evidence for the presence of barren plateaus for the lattice under discussion.

For all the RI cases, the obtained energy did not deviate too much from $E=0$, which is a poor initial energy for an optimizer to start. We also plot the energies obtained after evolution on the N\'eel state and the ground state energy for comparison. The results show that initial energies obtained from random parameters are far away from the ground state energy, and assuming there are no local minima between the initial point and the global minimum, it would take a significantly large number of iterations as compared to starting from the N\'eel state to approximate the ground state energy. Thus, the N\'eel state is an efficient initial state for this example.

\section{Optimization progress \label{ap2}}
\begin{figure}
	\caption{Optimization progress curves showing the lowest energy at each iteration for different lattice sizes (see legend). \label{figap1}}
\includegraphics[scale = 1.1]{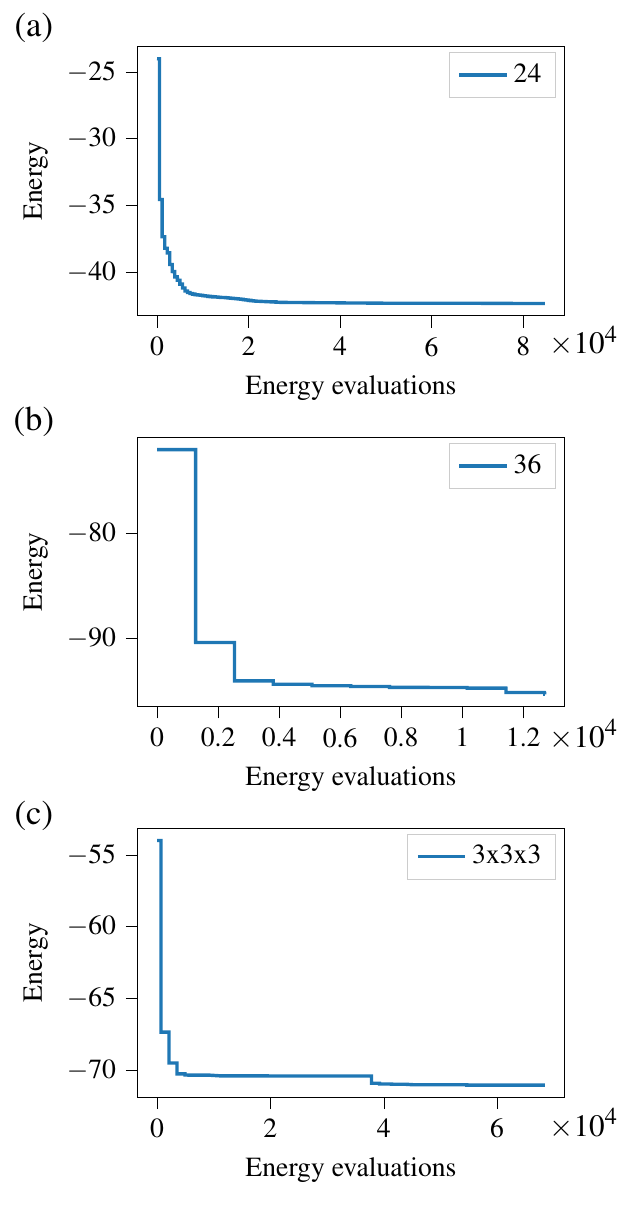}
\end{figure}

We show the energy progress at each step of the optimization for selected lattices. We plot the energy at each step of the optimization curve for $24$-qubit one-dimensional ring, a $6\times 6$ ($36$-qubit) square lattice, and a $3\times 3 \times 3$ ($27$-qubit) three-dimensional lattice, as shown in Fig.~\ref{figap1}. Initially, we observe a quick drop in the energy across all lattices, including those not shown here. As the VQE reaches a local minimum, the gradients become smaller and so does the relative decrease in the energy. This is reflected by the flat part of the curve towards the right of each plot. The step-like structure reflects the working of quasi-Newton optimization algorithms. 

As observed for the  $3\times 3 \times 3$ case,  after a significant initial drop, the drop in energy slows down after $10^4$ energy evaluations. This is explained by the fact that the optimizer gets trapped in a local minimum. However, by using the evolution heuristic, escape from the local minimum becomes possible (around $4\times 10^4$ evaluations), which shows a relatively large drop in the energy again. 

In Sec.~\ref{secqa}, we focus on comparing the time required to obtain the same optimization curves using an emulator and an ideal quantum computer. Fig.~\ref{figap1} shows that, on the one hand, obtaining the ground state requires a large number of energy evaluations, and on the other hand, the most considerable change in the energy drop occurs during the first few iterations. If the accuracy of the final energy is not important, the energy obtained within the first few iterations could also be used for comparing the computational times.

\end{appendices}

\bibliography{example}  

\begin{thebibliography}{85}
\bibitem [{not()}]{note1}%
  \BibitemOpen
  \href@noop {} {}\bibinfo {howpublished} {In principle, it is possible that
  the qubit with the largest number of gates takes less computational time than
  another qubit due to the different amounts of time required for each gate.
  The governing principle is to find the number of gates on the qubit that
  takes the longest to finish.}\BibitemShut {Stop}%
\end{thebibliography}

\end{document}